# Modulation of High-Energy Particles and the Heliospheric Current Sheet Tilts throughout 1976-2014

## M. A. El-Borie, S. A. Gwaily and A. A. Bishara

Physics, Department, Faculty of Science, Alexandria University, Egypt

## Abstract

Cosmic ray intensities (CRIs) recorded by sixteen neutron monitors have been used to study its dependence on the tilt angles (TA) of the heliospheric current sheet (HCS) during period 1976-2014, which covers three solar activity cycles 21, 22 and 23. The median primary rigidity covers the range 16-33 GV. Our results have indicated that the CRIs are directly sensitive to, and organized by, the interplanetary magnetic field (IMF) and its neutral sheet inclinations. The observed differences in the sensitivity of cosmic ray intensity to changes in the neutral sheet tilt angles before and after the reversal of interplanetary magnetic field polarity have been studied. Much stronger intensity-tilt angle correlation was found when the solar magnetic field in the North Polar Region was directed inward than it was outward. The rigidity dependence of sensitivities of cosmic rays differs according to the IMF polarity, for the periods 1981-1988 and 2001-2008 (qA < 0) it was $R^{-1.00}$ and $R^{-1.48}$ respectively, while for the 1991-1998 epoch (qA > 0) it was $R^{-1.35}$.

Furthermore, a moderate correlation was found between the TA of HCS and solar wind speeds (SWS) in the ecliptic plane. Also, a strong correlation was observed between sunspot numbers and tilt angles of neutral sheet, during the last three solar cycles 21, 22 and 23. These correlations are dependent on the essence of each SA cycle, whereas for the even solar cycle (22) the degree of the correlation is much steeper than that of the odd solar cycles (21 & 23). A somewhat low relation is found between SW density and TA; as well as, inclinations of HCS are linked with solar Temperature and standard deviation of IMF by weak correlations. It was remarkable that the sigma of IMF and alpha relation is not dependent on the sort of the solar cycle (odd or even) counter to the other four studied solar and interplanetary indices.

Hysteresis loops between TA and CRIs have been examined during three solar activity cycles 21, 22 and 23. Marked differences during the two odd and the even solar cycles, as well as during different polarity states of the solar magnetic field (qA > 0 and qA < 0) have been found. We observe variations in the features of modulation loops obtained using one, three, six and twelve solar rotation averaged data. A consider differences in time lags during qA > 0 and qA < 0 polarity states of the heliosphere have been observed. We also found that the cosmic ray intensity decreases at much faster rate with increase of tilt angle during qA < 0 than qA > 0, indicating stronger response to the tilt angle changes during qA < 0. Our results are discussed in the light of 3D modulation models including the gradient, curvature drifts and the tilt of the heliospheric current sheet.

# 1. Introduction

The sector structure observed in the interplanetary magnetic field (IMF) may be interpreted in terms of a warped equatorial current sheet in the heliosphere (e.g., Wilcox et al., 1980; Smith et al., 1978). Hence, the heliospheric current sheet, HCS, is the boundary encircling the Sun that separates oppositely directed heliospheric magnetic fields that originate on the Sun and are open (only one end is attached to the Sun). These fields are closely associated with the Sun's dipole magnetic field and have opposite magnetic polarities, e.g., outward (positive) in the north and inward (negative) in the south. The HCS is unique and represents the magnetic equator of the global heliosphere.

One of the most important parameters characterizing the structure of heliospheric magnetic field is the tilt angle of the heliospheric current sheet ($\alpha$), which corresponds to the heliomagnetic equator. The magnetic axis of the Sun is tilted with respect to rotational axis and, together with the Sun's rotation and radially expanding solar wind, the sheet forms a complicated 3D-structure. The tilt angle has been determined at the Wilcox solar observatory since 1976. The tilt angle is important for the large scale magnetic field and the solar dynamo, being related to the inclination of the Sun's magnetic dipole axis. The HCS tilt angle is also a key parameter of galactic cosmic ray modulation in the heliosphere. The relation between the HCS tilt angles and variations of the cosmic ray intensity has been intensively studied both theoretically (e.g., Fisk et al., 1998; Potgieter and Ferreira, 2001) and empirically (e.g., Belov, 2000; El-Borie, 2001). As HCS divides the solar magnetic field into hemispheres of opposite polarity, the ≈ 11-year period when it is directed outwards in the northern hemisphere has become known as qA > 0 epochs, such as during the 1970s and 1990s, while the 1980s and the period 2002-2014 are known as qA < 0 cycles. The HCS has a wavy structure, parameterized by using its tilt angle $\alpha$ (Hoeksema, 1992) and is well correlated to solar activity. The waviness of the HCS plays an important role in CR modulation (Smith, 2001). It is still the best proxy for solar activity from this point of view. It is widely used in numerical modeling.

The time-dependent drift model was used to illustrate the modulation effect of a simulated wavy neutral sheet which evolves with time and propagates outwards in the heliosphere at the solar wind speed. Studies by Swinson et al. (1990) have concluded that the asymmetries in solar activity (SA) between the northern and southern hemispheres of the Sun could influence the position of the neutral sheet source on the Sun. The source of the neutral sheet displaced away from the solar hemisphere which had higher level of solar activity as expressed by the sunspot number. This in turn leads to the Earth spending more time either above or below the neutral sheet, resulting in an asymmetry in the frequency of toward or away polarity of the IMF near the Earth.

Studies of El-Borie (1999; 2001) showed that the HCS crossings the Earth were associated with different effects on cosmic rays dependent on the solar polarity states with differences in both cosmic ray density and time profile. The cosmic ray intensities at Earth, observed by neutron monitors, were higher at the time of the sector boundary passing the Earth than either side. Negative cosmic ray gradients relative to the location of the neutral sheet during both solar magnetic phases, were observed. The high speed solar wind streams were not associated with the current sheet arrival on the Earth (El-Borie et al., 1996; 1997; 1998). Studies of various plasma parameters, including solar wind speed, density, temperature, and composition (ratio of alpha particle to proton densities), showed a close correlation with the current sheet (Smith, 2001). Nearly strong correlations (with magnitude ≈ 0.33-0.45%) were seen between α and the averages of solar wind speed during different periods (El-Borie, 1999).

Alania et al. (1995) displayed that the energy hysteresis of the galactic cosmic ray intensity ($R_m$ = 1- 6 GV) was not seen during the whole 11-year cycle but only during 5-6 years close to solar activity maximum. The hysteresis began with a typical increase of low energy cosmic rays followed by an increase of high rigidity cosmic rays. The energy spectrum of the cosmic ray isotropic variation was soft in solar maximum activity periods and it was hard in the minimum ones.

The purpose of this work is to examine the correlations of heliospheric current sheet tilts with cosmic ray intensities of different energies, recorded by sixteen neutron monitors, by considering evolution of the heliospheric current sheet during the period 1976–2014, which includes the entire three solar activity cycles 21, 22 and 23. The sensitivities of cosmic ray intensity to changes in the neutral sheet tilt angles before and after the reversals of the polarity of interplanetary magnetic field would be investigated. We also compare the variations of the energy spectrum with the waviness changes, during the two epochs of the interplanetary magnetic field phase (qA>0 and qA<0). As the tilt of the heliospheric current sheet is thought to play central role in drift models of galactic cosmic ray modulation, we attempt to study the time lag, hysteresis effect, area of the hysteresis loop, cosmic ray response to changes in tilt angle during different solar and magnetic conditions. As well, this work can be improved by including the rigidity dependent effect on galactic cosmic ray (GCR) modulation phase that may result from a time-dependent inclination angles that gradually increases (decreases) during the descending (ascending) phase from the outer heliosphere to the Sun.

In addition, time series of the HCS tilt angle (α), count rates of cosmic rays (CRs) (observed by sixteen neutron monitors), sunspot number ($R_Z$), as well as interplanetary magnetic field parameters (solar wind speed, ion density, plasma temperature and standard deviation of IMF vector) ought to be compared during different phases of the 21, 22 and 23 solar cycles. With new and recent data, general trends and solar cycle variations are evaluated and compared with published results. Points of correspondence and differences are noticed.

Furthermore, we seek to analyze correlations between the HCS maximal latitudes and solar activity parameters like sunspot number, solar wind speed, solar wind density, solar wind temperature and root mean square of heliospheric magnetic field during the three phases of solar activity. Tilt angle (TA, waviness of heliospheric neutral current sheet) and solar/ interplanetary parameters have been considered as solar activity (SA) parameters to study the long-term modulation of CRI for the period of 1976 to 2014. Here, "Running cross correlation analyses" has been applied to study the relationship between tilt angle of HCS (alpha) and, CRI observed at Oulu NM, sunspot number ($R_Z$), solar wind speed (SWS) and solar ion density (n) during the whole investigation period.

## 2. Data collection

In the present work, tilt angles (i.e., α, the waviness of heliospheric neutral current sheet) has been considered as a solar activity (SA) parameter to study the long-term modulations of cosmic ray intensity (CRI) during the period from 1976 to 2014. To study the average behavior of cosmic ray intensity, mesurements of α were provided via the Wilcox Solar Observatory (WSO). WSO`s official site (http://wso.stanford.edu/Tilts.html) gives the maximum extent in latitude reached by the computed heliospheric current sheet (HCS) from 1642 to about 2146 Carrington rotation number (CN). Values of HCS tilts are provided for the maximum northern and southern extent of the HCS as well as the average for both classic and radial models. The maximum latitudinal extent of the values above 70 degrees should be considered lower limits as the model is not computed over 70 degrees. The two model HCS's generally track each other, but with a higher source surface radius (R = 2.5 or 3.25 solar radius, $R_S$), the radial model current sheet shown here is flatter and therefore usually has a lower maximum extent. The classic model (R = 2.5 $R_S$) results tabulated here probably gives a better estimate and so it is used in this work (Potgieter and Ferreira; 2001).

One hour-averaged data corrected for atmospheric pressure and corrected for efficiency changes with barometric coefficient variable with solar cycle for 16 NMs, were used and analyzed for the period from May 1976 to January 2014. This period covers three solar cycles (21, 22 and 23). The world network of ground based NMs (NMDB; http://www.nmdb.eu) provide very stable real time intensities of cosmic rays with different rigidities. **Table 1** illustrates characteristics of the 16 selected detectors, which are geographic coordinates, altitude from the sea level, cutoff ($R_C$) and the median primary ($R_m$) rigidities.

The website ( http://omniweb.gsfc.nasa.gov/ ) provided us with data set which is primarily a 1976-to-2014 compilation of hourly-averaged, near-Earth solar wind, magnetic field, sunspot number and plasma parameters (temperature and density) data from several spacecrafts in geocentric or interspersed near 1 AU.

## 3. Data treatment and analysis

Time resolution of available α-points is daily hourly-averaged every ~ 27 day (i.e. some point at some hour at some day, observed every 27 day, and so on). This is the type of resolution that we follow and use in our study. Neither gaps nor zeros were seen between tilt angles so it is considered as a reliable one. It was observed that the α-values changed periodically with time, whereas the minimum values 13.2º, 8.4º, 8.9º and 10.1º were recorded at (minima solar activity) 8/1976, 2/1987, 2/1996 and 10/2009 respectively and the maximum ones 75º, 75º, 74.9º, and 74.6º were recorded at maxima solar activity; 5/1979, 3/1990, 6/1999, and 6/2012, respectively.

We prepared a computer program which was written in the simple programming language, JAVA. The Java program run every time on one of the NMs data of one-hour averaged as input and output count rates with 27-day averaged. Following, some gaps, missed counts and zeros were detected at few neutron monitors and in order to improve the reliability of the recorded data and to prevent measurement breakdowns, two independent processes were performed on data. First, all bad data was erased and replaced by gaps. Second, for few missing counts, Spline interpolation action has been performed on whole neutron monitors counts automatically to hold existing gaps with good predicted ones. The same procedures applied on CRIs are also used to treat hourly data of Omni web.

**Table 1: Characteristics of 16 NMs stations used in the present work**

| No. | Station Name | Symbol | Latitude (degree) | Longitude (degree) | Altitude (m.a.s.l) | $R_C$ (GV) | $R_m$ (GV) | Location |
|---|---|---|---|---|---|---|---|---|
| 1 | Thule | THU | 76.6 | -68.8 | 260 | 0 | 16 | Greenland |
| 2 | McMurdo | MCM | -77.95 | 166.6 | 48 | 0 | 16 | Antarctica |
| 3 | Inuvik | INVK | 68.35 | -133.72 | 21 | 0.17 | 16 | Canada |
| 4 | Apatity | APT | 67.55 | 33.33 | 177 | 0.65 | 16 | Russia |
| 5 | Oulu | OUL | 65.05 | 25.47 | 15 | 0.8 | 16 | Finland |
| 6 | Calgary | CAL | 51.08 | -114.13 | 1128 | 1.09 | 17 | Canada |
| 7 | Newark | NEWK | 39.7 | -75.7 | 50 | 2.09 | 18 | USA |
| 8 | Kiel | KIEL | 54.30 | 10.10 | 54 | 2.36 | 18 | Germany |
| 9 | Moscow | MOS | 55.47 | 37.32 | 200 | 2.43 | 18 | Russia |
| 10 | Hermanus | HER | -34.42 | 19.23 | 26 | 4.44 | 20 | South Africa |
| 11 | Rome | ROME | 41.9 | 12.5 | 60 | 6.32 | 22 | Italy |
| 12 | Tbilisi | TBL | 41.72 | 44.8 | 510 | 6.73 | 23 | Georgia |
| 13 | Potchefstroom | Potch | -26.68 | 27.1 | 1351 | 6.98 | 25 | South Africa |
| 14 | Tsumeb | TSM | -19.2 | 17.58 | 1240 | 9.12 | 27 | Namibia |
| 15/16 | Huancayo/Haleakala | HUN/HAL | -12.03/20.72 | -75.33/-156.28 | 3400/3052 | 12.92/12.91 | 33 | Peru/Hawaii |

## 4. Results and Discussion

### 4.1 The HCS Tilts, CR Intensities and Solar and Inter-planetary Parameters during Three Solar Cycles.

The Sun is the dominant variable source, which controls the structure of the heliosphere, the modulation of cosmic rays through the level of solar activity, the tilt angles of the heliospheric current sheet, the solar wind speeds, and the strength and turbulence of the interplanetary magnetic field (McDonald et al., 2010). It will be interesting to compare the nature and magnitude of variabilities in solar activity (*e.g.*, sunspot number), solar wind speed, interplanetary magnetic field strength and its fluctuations with the tilts of the heliospheric current sheet during similar phases (*e.g.*, minimum & maximum) of different solar cycles.

Studies of long-term variation of the tilts of heliospheric current sheet, cosmic ray intensities and the solar activity parameters represent a general picture of negative and positive relationships, and provide a broad qualitative idea about the relationships between them. Figure 1 represents a time series study from May 1976 to January 2012 over 478-Carrington solar rotations (from CN 1642 to CN 2119) of 27day-averaged tilt angles of heliospheric current sheet α (plot a), averaged cosmic ray intensities (CRI) measured by Oulu neutron monitor for which the cutoff cosmic ray rigidity is ≈ 0.8 GV (plot b), smoothed sunspot number $R_Z$ (plot c), smoothed solar wind speed SWS (plot d), smoothed root mean square of interplanetary magnetic field vector $\sigma_B$ (plot e), smoothed solar wind plasma density n (plot f) and smoothed SW Plasma temperature T (plot g). The periods between the three thin dashed vertical lines, shown in the Figure around 1980, 1990 and 2000 correspond to the solar polar magnetic field reversals (switch between the magnetic field polarity states qA > 0 (IMF points outward and away from the Sun) and qA < 0 (IMF points inward and toward the Sun).

Figure 1a displays the time series of tilt angles of the neutral sheet over the 37-years interval; from the plot it is clear that:

- ❖ At minimum solar activity years 1976 and 1977 (with CNs from 1642 to 1663), the inclinations were found to have a value less than 37°. Hence, the inclinations have gradually increased till the solar maximum years 1979-1980 (CNs from 1677 to 1703) where 16 tilt greater than 70° were found. Low inclination angles initiated to appear again at the beginning of 1985 (CN 1758) and they persevered through the minimum solar activity of years 1986-87 (CNs from 1771 to 1797). Towards February 1987 (CN 1786), the HCS had a very low tilt, as expected. From May 1988 (CN 1802), tilt angles started again to rise up to the solar maximum 1989-1990 (CNs from 1811 to 1837). A third and forth minimum activity of low inclination values existed at 1996-1997 (CNs from 1905 to 1931) and 2008-2009 (from CN 2066 to CN 2091), respectively, noting that the next maximum activity was during the years 1999 and 2000 (CNs from 1945 to 1971).
- ❖ The α-modulation during both the solar minimum periods 1976-1977 and 1986-1987 was almost the same unlike the 2008-2009 period which posses a belated and wide solar minimum. This remarkable variation originated from distinct beginnings and ends of solar cycles 21, 22 and 23. Traditionally, the solar cycle extends amongst two successive minimums. So, according to plot 1a, the three solar cycles 21, 22 and 23 are shown. Through the above-mentioned solar minimum periods, the current sheet owned a lower tilt angle and the Earth spent less time away from the neutral sheet and underwent to greater cosmic ray intensities. As well, it can be seen from plot 1a that there are a few distinguished wanes in the current sheet tilt angles. Indicating to the three deeply decreases in α, the first from 61.3° to 31.2° occurred in September 1982 to December 1983 (CNs from 1726 to 1743), the second from 66.7° to 23.9° been in February 1991 to May 1992 (CNs from 1839 to 1856) and the third decrease from 66.3° to 40° at the beginning of 2001 (CN 1974). Thus, the solar cycle variation of the neutral sheet tilt angles is approved by high values around the maxima solar activity beside dunks to lower values occurs near the minima solar activity.

From the two plots a and b of Figure 1, it is noted an inverse correlation between cosmic ray intensity and the tilts of HCS over the entire period. In plot b, there are noteworthy CR intensity drops in May 1978 and April 1979 (close to the first solar maximum), in May and December 1989 (near the second solar peak) and in August 1998 and June 2000 (around the third peak). These depressions in the count rats of CR pursued excursions and revolutions in the tilts of the HCS. Plus, another five diminutions happened respectively in, November 1982 (CN 1729), May 1984 (CN 1749), February 1986 (CN 1772), July 1991 (CN 1845) and finally April 2001 (CN 1975). The last CR events are owing to some solar phenomena and are not linked to any changes occurred in the inclinations of the current sheet, in agreement with previous spotted correlations between alpha and CRI by (Hundhausen et al., 1980; Saito et al., 1986; El-Borie, 2001 and Badruddin et al., 2007).

The level of solar activity is traditionally represents by sunspot numbers $R_Z$. A comparison of smoothed sunspot numbers during the declining to minimum and rising till maximum phases of three solar cycles 21, 22, and 23 is shown in plot 1c. To compare the three minima, in particular, first point in this plot corresponds to the end of minimum solar activity cycle 20 (after which the solar activity starts increasing for the next solar cycle). As compared to the previous two solar cycles, the decline of solar activity in cycle 23 toward minimum is the longest and the weakest one. Sunspot number and the tilt of HCS relate to the solar dynamo differently. Furthermore, both reflect the structure and evolution of the solar magnetic field. One can see that the tilt of HCS changes systematically over the observed solar cycles and the variations of the HCS inclinations are in good agreement with the sunspot number variations during the considered period. Suess et al. (1993) showed that the tilts for solar cycle 21 were overlayed onto solar cycle 22 using an autocorrelation between the cycles to guide the overlay. It has been obtained the maximum autocorrelation between the tilt angle data for cycles 21 and 22 for a time shift of 10.5 years. Thus the HCS tilts could also, like other related activity factors, evolve in non-uniform fashion from cycle to cycle.

The character of the solar wind and heliospheric magnetic field undergoes cyclic changes that accompany the 11-year (or 22-year) solar cycle. It has been clear by (El-Borie et al., 1997; 1998) that the high speed solar wind streams were not associated with the HCS crossing our Earth. The smoothed variations in the solar wind speeds SWSs in the same period are plotted in plot 1d. Although the smoothed variations in the solar wind speed do not strictly follow the sunspot number variations, the speeds are slowest during the latest minimum of cycle 23, in comparison to the previous two minima. As well as, it has been found that there is a great agreement between the tilt angles and the smoothed solar wind speeds. The solar wind speed increases significantly with distance from the current sheet, whereas, when the HCS is inclined and be wavy, the spacecraft (which are in the ecliptic plane) spends a greater fraction of time away from the current sheet and so experiences higher SWS (Saito et al., 1986).

When similar phases of three solar cycles (21, 22 and 23) are compared, the smoothed plot (e) of standard deviations of IMF vector $\sigma_B$ (which is considered as a measure of fluctuations in the interplanetary magnetic field), the smoothed plot (f) of solar wind plasma densities and finally the smoothed plot (g) of solar wind plasma temperatures, show a weak response to changes in the tilts of HCS. The last three time dependence diagrams (e, f & g) show small variations of pre-mentioned parameters which change very slowly with time and been weakest at the minimum of the solar cycle 23. These parameters, and many more, are summarized during the deep minimum of solar cycle 23 by (Jian et al., 2011). However, in contrast to the sunspot number, the solar wind velocity and magnetic field, the tilt of the HCS is not the smallest during cycle 23, but larger than that during the previous two minima of cycles 21 and 22.

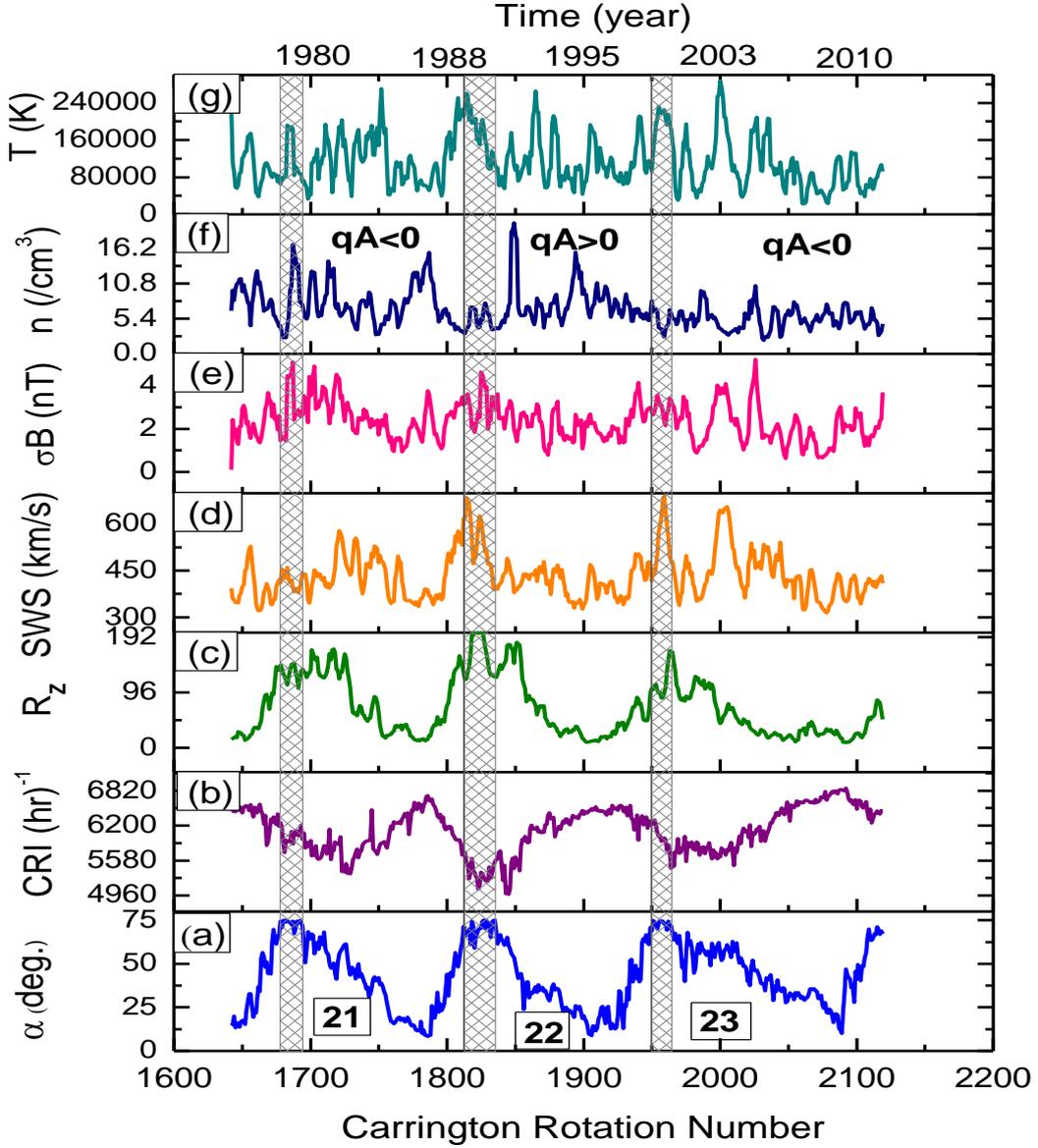

Figure 1: Time profiles of the heliospheric current sheet (HCS) tilt angles α (a), the cosmic-ray intensity (CRI) from Oulu neutron monitor station (b), the smoothed sunspot number $R_Z$ (c), the smoothed solar wind speed sws (d), smoothed rms of IMF Vector $\sigma_B$ (e), smoothed ion density n (f) and smoothed sw plasma temperature T (g) for each Carrington rotation from 1642 to 2119 (down scale), for the period 1976 – 2012 (top scale). The three thin dashed vertical lines separate periods of different IMF polarities (qA < 0 or qA > 0).

## 4.2 Correlation of the HCS Tilts to Galactic CRIs of Various Rigidities

Since the solar activity is not directly related with cosmic ray intensities, but the heliospheric current sheet structure and its tilt angles certainly provide a good solar physical view to explain the long-term galactic cosmic ray modulations in the heliosphere. To understand whether the CR intensities access route to the inner heliosphere is the main cause that makes this difference in responsiveness of the tilt of the HCS, we present, in this section, regression analysis between tilt angles of HCS and CRIs by separating the selected period (1976-2012) into three groups, which are pre- and post- the IMF reversal (when the IMF configuration changed) in 1989/1990 and post- the IMF reversal in 1999/2000, to cover the solar magnetic polarity states (qA>0 & qA<0). Times of solar polarity reversals and very high solar activities (1979/1980, 1989/1990 and 1999/2000) have been omitted due to the complex structure of HCS. During these years the morphology of the HCS was quite complex and no regular pattern was observed (Hoeksema et al., 1982; 1983; 1989). Between polarity reversals, the neutral sheets had relatively "well-behaved" and consistent shapes. Therefore, α-values ≥ 70° have been removed, since it is difficult to study the values of α above 70° (Hoeksema, 1989).

Figures 2–4 display the scatter plots between the tilt angles, α, of heliospheric current sheet (HCS) and cosmic ray intensities (CRIs) through a rigidity range from 16 to 33 GV. The anatomized periods and Carrington rotation numbers are noted in each plot. The best-fit line to the data, the magnitude of the sensitivity of CRIs to change with tilt angles, a, the regression coefficient, r, for each fitted curve, are shown. The cutoff rigidity, $R_O$, for each neutron monitor used is also indicated.

In agreement with earlier observations (El-Borie, 1999, 2001; Umakant Sharma et al., 2005; Badruddin et al., 2007), there is a negative and high correlation between CRIs and the inclinations of HCS. The magnitude of the correlation depends on the solar magnetic polarity states (qA < 0 or qA > 0). The inverse correlation is more pronounced when the solar magnetic phase was negative (on average ≈ - 0.838 for 1981-1988 and ≈ - 0.892 for 2001-2008) than when it is positive (≈ - 0.757 for the 1991-1998 epoch). This implies that the inverse correlation is a rigidity and IMF polarity–dependence. We notice that there is a considerable scatter in the data; part of it is probably due to the uncertainties in the values of tilt angles. Furthermore, there are more scatters of the α-CRIs plots during the period of qA > 0 in coincidence with α ≥ 45° than the other periods of qA < 0.

The slope of the best-fit straight line, which is a measure of the CR sensitivity to the variable tilts, for the period 2001-2008 (epoch of negative magnetic field orientation) is much steeper than the period 1991-1998 (epoch of positive MF) which in turn has quasi a lower slope of fitted line than that of the 1981-1988 interval (years of positive polarity). In addition, the CRI is more robustly affected by changes in the tilt of the current sheet during the period of negative half cycle of magnetic polarity (e.g., 4771 counts/hr/deg for McMurdo at 2001-2008 interval), whilst it is 2957 counts/hr/deg during the period of positive half cycle, a half times larger. The CRI observed at Haleakala NM which is close to magnetic equator, we observed less sensitivity to the changes in α (≈ 177 counts/hr/deg) during the qA > 0 epoch (1991-1998). The observed sensitivity and the inverse correlations (detected at Haleakala of $R_O$ = 12.91 GV) were much larger for qA < 0 (at 2001-2008) than for qA > 0 (at 1991-1998). Accordingly, for the qA > 0 epoch, we propose that the cosmic ray modulations by the HCS tilt angles became weaker with increasing particle energy. Our findings are consistent with the predictions of the gradient drift model. The sensitivity of CRs to the changing angle of the current sheet was smaller than when the field above the current sheet was outward (positive IMF status) (El-Borie, 2001).

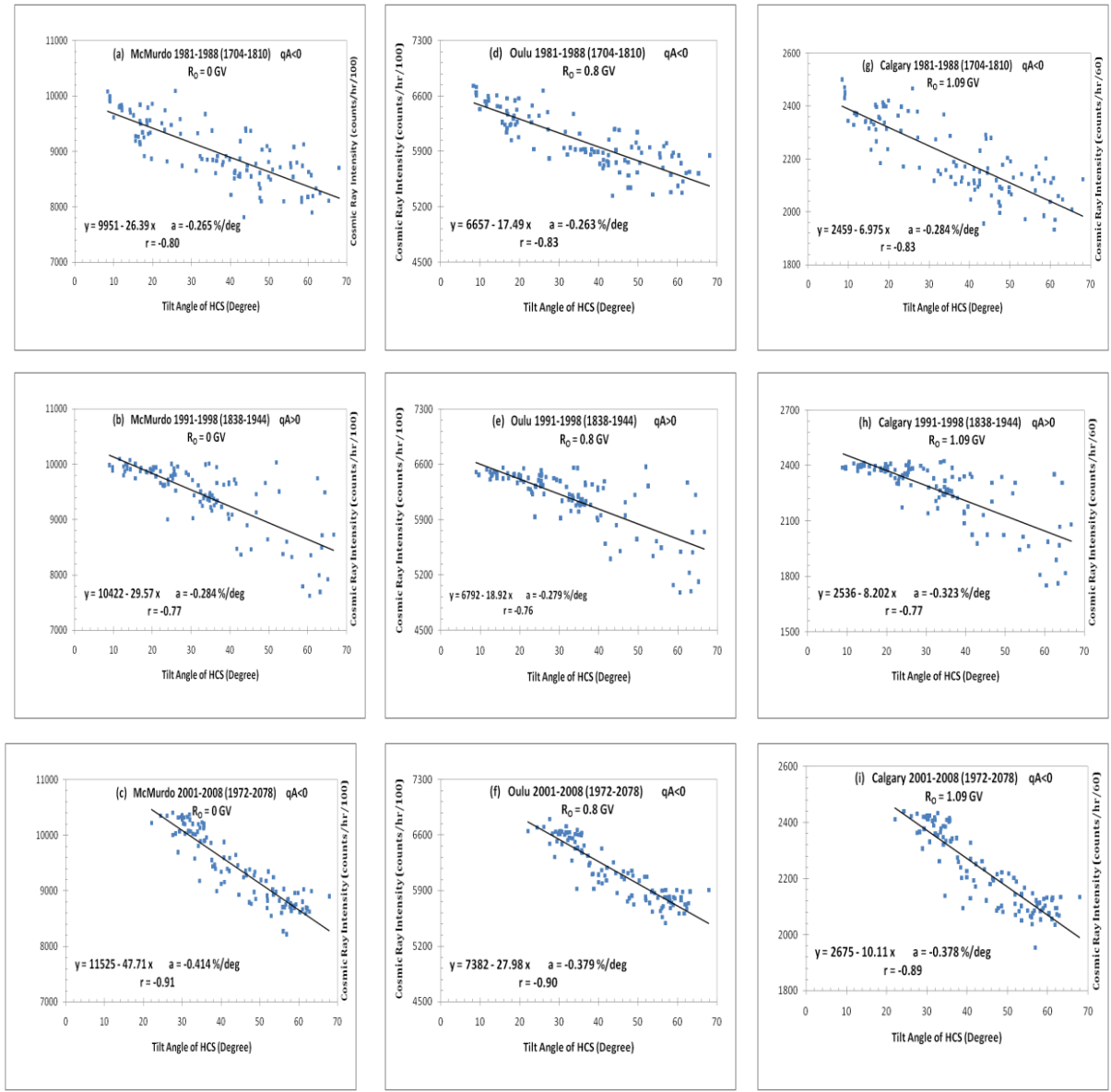

Figure 2: Scatter plots and best fit curves between tilt angles of HCS and GCR intensities observed at McMurdo (MCM), in left panels, Oulu (OUL), in middle panels, and Calgary (CAL), in right panels, during three periods 1981-1988 (IMF polarity qA < 0), 1991-1998 (IMF switched polarity to qA > 0) and 2001-2008 (IMF turned polarity again to qA < 0). The eqution of fitted line, the magnitude of sensitivity of CRI to changing in tilt angle (a), the correlation coefficient (r) and the cutoff rigidity for each NM ($R_c$), are noted. Carrington rotation numbers are displayed in each plot.

The ratio of the average slopes of the regression lines for qA < 0 (2001-2008 duration) to qA > 0 (1991-1998 epoch) for all used detectors ranges between 1.15 at Haleakala, and 1.63 at Potch. So, the CRIs are affected by the tilts when the positive CR particles reach the inner heliospheric by drifting in along the neutral sheet than out along it. The mean ratio of the regression coefficient, r (qA < 0, 2001-2008) / r (qA > 0, 1991-1998), for all NMs, is found to be ≈ 1.18. Therefore, it is obvious that the tilts α can be used as a meaningful and significant solar parameter, especially in the case of cosmic ray modulation. Investigations by Badruddin et al. (2007) showed a correlation between Oulu/NM and Climax/NM intensities and the tilt angle for 1, 3, 6 and 12 rotation averaged data during the period 1976-2006. The slopes of the regression lines for the 2001-2006 period (qA < 0) was almost a factor of two (average of 1.74 & 2.23) larger than for the 1991-1999 period (qA > 0). While, the ratio of correlation coefficients, r, between epochs qA < 0 and qA > 0, extended between ≈ 1.08 and ≈ 1.51. Similar comparisons made for the 1981-1987 period (El-Borie, 1999; 2001); by using the Potchefstroom NM data yielded a slope of 0.17 % per degree. Our results are in good agreement with 0.175 % per degree for the period 1981-1988.

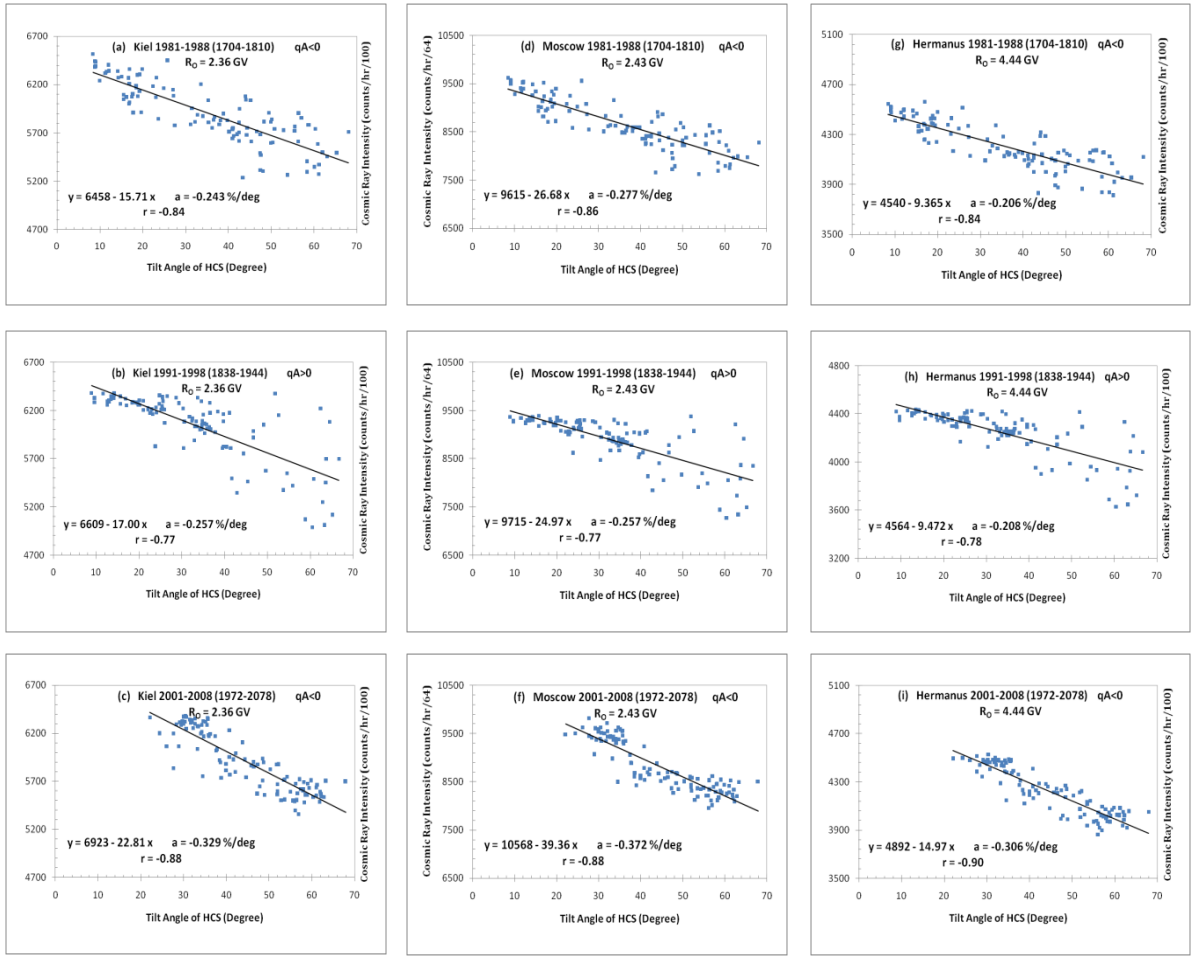

Figure 3: Same as Figure 2 for Kiel (KIEL), in left panels, Moscow (MOS), in middle panels, and Hermanus (HER), in right panels.

It is obviously seen from Figures 2-4, that when the north polar cap has a negative polarity, the cosmic ray diminution is more rapid than during the interval 1991-1998; the sensitivity of count rate on inclination is larger during 1981-1988 and much larger during 2001-2008. At Rome NM in Figure 4, the average count rate of cosmic ray decreases from 5042 to 4300 counts/hr/100 when α changes from 10° to 70° during 2001-2008 epoch (qA < 0). In contrast, the decrease is from 4767 to 4299 counts/hr/100 for the same α range for the period 1991-1998 (qA > 0). The dispersion of points is perceivable despite the fact that this is the best correlated set among all those used in this work.

Results of our analysis are in consistent with cosmic ray distribution in inner heliosphere in which cosmic ray intensity of neutral sheet is highest, with intensity decreasing with distance laterally from the sheet, when the tilt angle α increases and warp in the neutral sheet becomes more pronounced, the Earth spends more time of greater distances from the neutral current sheet and therefore is exposed to a lower cosmic ray intensity.

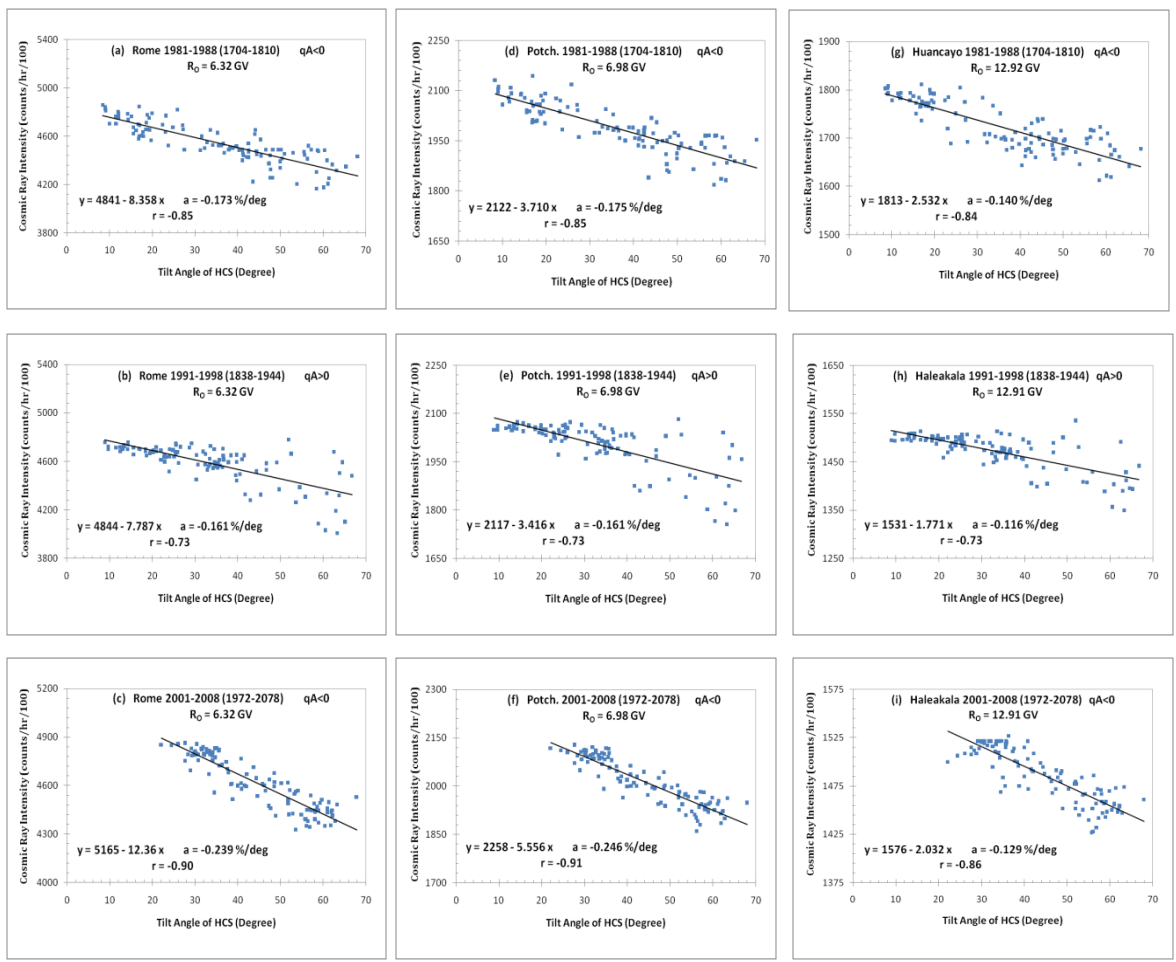

Figure 4: Same as Figure 2 for Rome (ROME), in left panels, Potchefstroom (Potch.), in middle panels, and Huancayo/Haleakala (HUN/HAL), in right panels.

Cosmic ray drift models depend on the three dimensional topology of interplanetary medium and vary according to solar activity. We know from the predictions of these models that gradient and curvature drift can cause cosmic rays to travel through vastly different regions of the heliosphere in consecutive solar cycles. When the northern heliospheric magnetic field is directed outward (qA > 0), particles with positive charge drift inward from the polar regions to the equatorial plane and out along the wavy neutral sheet. The waviness of the neutral sheet, which is indicated by the so called tilt angle α is then expected to play a relative unimportant role in modulation, leading to broad galactic cosmic ray maxima (Jokipii et al., 1981; 1983). When the northern field is directed inward (qA < 0), positive particles reach the inner heliosphere by drifting in along the neutral sheet and the waviness of the neutral sheet is therefore in this case expected to play a more important role in the modulation processes, leading to more peaked or triangular galactic cosmic ray maxima ( El-Borie, 1997; 1998 ).

## 3.3 Rigidity-Dependent Effect in the modulation of GCR Intensity

The rigidity dependence of modulation arises from the local and global GCR contributions. To explore this dependence, we used data obtained with a variety of detectors at sea level and mountain sites by balloons, satellites and space probes (see Table 2).

**Table 2: Slope of the regression lines between tilt angles and CRIs, a, for the selected detectors for different polarity states.**

| No. | NM Station | $R_m$ (GV) | - a [%/deg] | | |
|---|---|---|---|---|---|
| | | | (1981-1988) | ( 1991-1998) | (2001-2008) |
| 1 | Thule | 16 | 0.263 | 0.273 | 0.388 |
| 2 | McMurdo | 16 | 0.265 | 0.284 | 0.414 |
| 3 | Inuvik | 16 | 0.254 | 0.284 | 0.393 |
| 4 | Apatity | 16 | 0.263 | 0.286 | 0.399 |
| 5 | Oulu | 16 | 0.263 | 0.279 | 0.379 |
| 6 | Calgary | 17 | 0.284 | 0.323 | 0.378 |
| 7 | Newark | 18 | 0.246 | 0.261 | 0.345 |
| 8 | Kiel | 18 | 0.243 | 0.257 | 0.329 |
| 9 | Moscow | 18 | 0.277 | 0.257 | 0.372 |
| 10 | Hermanus | 20 | 0.206 | 0.208 | 0.306 |
| 11 | Rome | 22 | 0.173 | 0.161 | 0.239 |
| 12 | Tbilisi | 23 | 0.191 | 0.175 | 0.181 |
| 13 | Potchefstroom | 25 | 0.175 | 0.161 | 0.246 |
| 14 | Tsumeb | 27 | 0.146 | 0.137 | 0.203 |
| 15/16 | Huancayo/ Haleakala | 33 | 0.14 | 0.116 | 0.129 |

For such studies to be meaningful, it is important to have a clear understanding of the response characteristics of the detectors involved. We characterize them in terms of their median rigidity of response ($R_m$) to GCR spectrum; i.e. below it is 50 % of a detector counting rate (El-Borie; 2001). It is easily computed from the detector response function, derived from the latitude survey data at sea level and mountain altitudes. Data are available from several surveys undertaken by different research groups over several decades. Most of the surveys are carried out near solar minima for an understandable reason that minima are free of the transient events (solar cosmic rays, Forbush decreases, etc); also the intensity of lower rigidity GCRs is the largest of them. Lockwood and Webber (1996) studied the inverse rigidity dependence of modulation for cycles 21 and 22. The $R_m$ values computed are based on two assumptions. First, modulation ceases above 100 GV. Second, $R_m$ value depends upon the modulation function assumed.

Figure 5 displays the sensitivities of tilt angles of HCS (for $\alpha \leq 70°$) to CRIs and the median response rigidity of each detector during the three periods 1981-1988 (qA < 0), 1991-1998 (qA > 0) and 2001-2008 (qA < 0). The dashed, the dotted and the solid lines are the regression curves for the years before and after the IMF reversal in 1989/1990 and for the period 2001-2008, respectively. The cosmic ray sesitivities to the inclinations of current sheet, is rigidity dependent. The intensities varied with α for the magnetic half cycle in which the Sun`s polar cap field and the heliospheric magnetic field were positive or outward in the north. This phenomena is proved by expressing the rigidity dependence in a power-law form, as indicated in Figure 5.

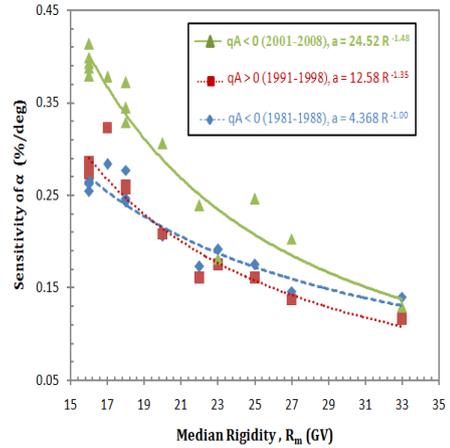

Figure 5: Sensitivities of $\alpha \leq 70°$ to GCRIs versus the median response rigidities. The green solid, the red dotted and the blue dashed lines represent the regression curves for the period 2001-2008, 1991-1998 and 1981-1988 respectively. The sensitivity-rigidity dependence follows a power-type law which it`s equations are mentioned.

A power fit for the data during the period 2001-2008 (qA < 0) and the 1981-1988 interval (qA < 0) yields $R^{-1.48}$ and $R^{-1.00}$ respectively, while a fit to the qA > 0 (1991-1998) values gives $R^{-1.35}$, indicating that the spectra of particles drift is harder in years after the IMF reversal 1999/2000 than before it. Subsequently, the reversal of the IMF polarity trends affects the rigidity reliance between the CRIs and the tilt angles of the HCS. Moreover, for particle median rigidities from 16 to 33 GV, the slope of the regression lines increase with decreasing energy; the rate of increasing is larger in the epoch 2001-2008. This shows a modulation for lower energy particles more obvious during the qA < 0 time. The drift model is applicable to NM observations at least near the onset phase of solar cycles 22 and 23 when the physical status of the heliosphere is much less complicated (Webber et al., 1988). The divergences in the time evolution of GCRIs from cycle to cycle can be attributed to the corresponding variations in tilt angles of HCS and to the differing sensitivity of GCR to α in both magnetic polarity states. The impact of the neutral sheet is however limited for particles entering from the solar regions and a difference in the rigidity dependence of the overall modulation, for the magnetic epochs 1981-1988, 1991-1998 and 2001-2008, consequently occurs.

To study the variation energy spectrum of CRIs caused by various HCS inclinations, we divided the tilt angles into three groups based on the computed HCS tilts for each Carrington rotation. Low tilt angles have values less than or equal to 30° (these values are labeled group I); group II has the values rang 30° ≤ α ≤ 50°, while high tilt angles have the range 50° ≤ α ≤ 70° (group III). Figure 6 (plots a to c) displays the observed sensitivities of CRIs to the tilt angles against the median response rigidities at six cosmic ray stations (Apatity- Calgary- Hermanus- Rome-Potch.-Tsumeb) for the two periods of IMF polarity state; 1991-1998 (qA > 0) and 2001-2008 (qA < 0) for the three groups, respectively. Figure 6a for group I is in agreement with the times of minimum solar activity, while group III meets the epochs near high solar activity records. Solid and dashed power-type fitted curves for qA > 0 and qA < 0 besides their equations are showed. In Figure 6a, the fit of the epoch qA > 0 gives $R^{-1.84}$, whereas the fit of the epoch qA < 0 grants $R^{-0.71}$. Consequently, the extent of the HCS waviness has more effectively evident role in cosmic ray modulations for qA > 0 than qA < 0 time. Furthermore, the rates of sensitivity of different energy particles (low and high) to alpha are higher during qA > 0 solar minimum periods (from mid 1994 to 1997; CNs from 1884 to 1931) than during the qA < 0 solar minimum. In the years 1994-1997 there are more (less) modulations at low (high) rigidities than in 2007's. Subsequently, the reversal of IMF polarities caused the computed spectra to be softer after the polarity change in 1999/2000.

Figure 6b presents the sensitivities of CRI to α group II which corresponds the period of solar activity during the ascending and descending intervals. Unlike group I, the rate of sensitivities of CRIs for qA > 0 ($R^{-1.41}$) is smaller than that for qA < 0 ($R^{-1.59}$). The variation spectrum became harder after the IMF reversal. Large modulations in CRIs are noted during the solar activity phases. The hysteresis effect for CRIs of different energy ranges has a common feature of cosmic ray modulation (Alania et al., 1995).

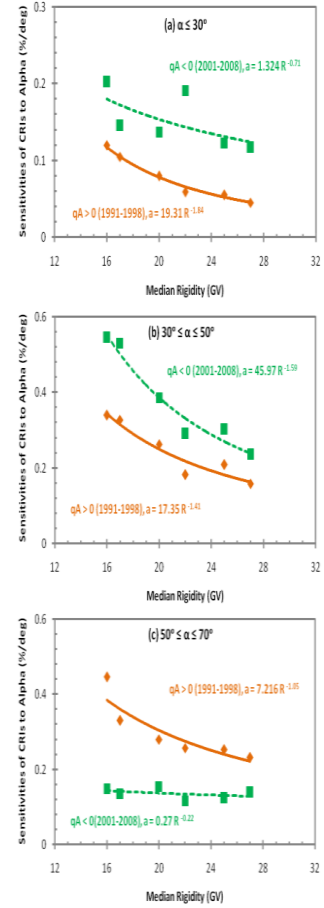

Figure 6: The sensitivities of CRIs to alpha versus median rigidity for six detectors for both IMF polarity epochs. Solid (dashed) line is the regression curve for the qA > 0 (qA < 0) epoch. The top, middle and bottom panels refer to α ≤ 30°, 30° ≥ α ≥ 50° and 50° ≥ α ≥ 70°, respectively.

Figure 6c shows the sensitivities of CRIs during and near the high solar activity years; 1991 (CNs from 1838 to 1849) for qA > 0 and from February 2001 to May 2004 (CNs from 1973 to 2016). α > 70° has been omitted. There are variances in the sensitivities and energy spectrum between both IMF epochs. The qA > 0 duration gives an energy exponent -1.05, while it gives -0.22 for qA < 0. So, the energy spectrum of cosmic ray modulations, caused by HCS tilts, has different behaviors during minimum and maximum solar activity periods as well IMF polarity epochs.

## 4.4 Correlations between HCS tilts and Solar/Interplanetary Magnetic Field parameters

The heliospheric neutral sheet is presumed to propagate from the Sun at solar wind speed up to the outer boundary of heliosphere. It is noted that, near the sector boundaries of current sheet, the solar wind speeds are permanently low and the high speed solar wind streams are not associated with crossing of the HCS at the Earth (Hakamada et al., 1984; El-Borie et al., 1997; 1998). Moreover, variations in solar wind speed and IMF magnitude, near crossings of neutral sheet, are linked to the distance of the Earth from the sheet (El-Borie et al., 1997; 1998). For investigating the correlation between the tilt angles of HCS and SWS for the considered period, we divided the data into three sets; June 1976- September1986, September 1986- May 1996 and May 1996- January 2008, corresponding to the three solar cycles 21, 22 and 23, respectively. These relations are shown in Figure 7. We presented the results of the linear regression analyses as the best-fit straight line, the output linear equation and the correlation coefficient, r. Carrington solar rotation numbers are displayed between brackets beside the studied period.

In general, positive and moderate correlations are observed between tilt angle and the SW plasma speed for the specified three epochs. A small but significant correlation coefficient (r = 0.131) for the period 1976-1986, is found. Also, the period 1996-2008 has a fairly small correlation coefficient (r = 0.339), while a larger coefficient (r = 0.499) is noted for the period 1986-1996. Moreover, the sensitivities of α to SWS recorded values of 0.622, 2.695 and 2.368 for solar cycles 21, 22 and 23, respectively. Hence, during the solar cycle 22, the slope of the fitted line is much steeper than the slopes of the other two solar cycles 21 & 23. The slope of the fitted line for solar cycle 22 is about four times than that for cycle 21.

In other words, the correlation between the tilts of HCS and the average solar wind speed is rather dependent on the essence of each solar cycle where the degree of the correlation in the even cycle (22) is almost twice the average of the magnitudes of the correlations in the odd ones (21 & 23). Newkirk et al. (1985) reported minimum SWS of ≈ 400 km/s at the HCS, with the speed rising to a mean speed of ≈ 600 km/s at angular distance from the current sheet of 20° to 40° during the years of low and intermediate solar activity. In a manner similar, El-Borie (1999) determined that the SWS rising (≈ from 380 km/s to 460 km/s) with a wide range of α (≈ from 10° to 65°) for the periods 1977-78 & 1987-89 on average, whilst for the period of descending phase of solar cycle 21 (1981-84), the speeds decreased ( ≈ from 520 km/s to 420 km/s with increasing alpha (≈ from 30° to 60°). From figure 7 (first plots), the SWS recorded a minimum value (≈ 300 km/s) exceeding the mean value (≈ 600 km/s) when the tilt angles of the HCS varied from ≈ 10° to ≈ 40° during the years of low and intermediate solar activity for the solar cycles 21, 22 and 23.

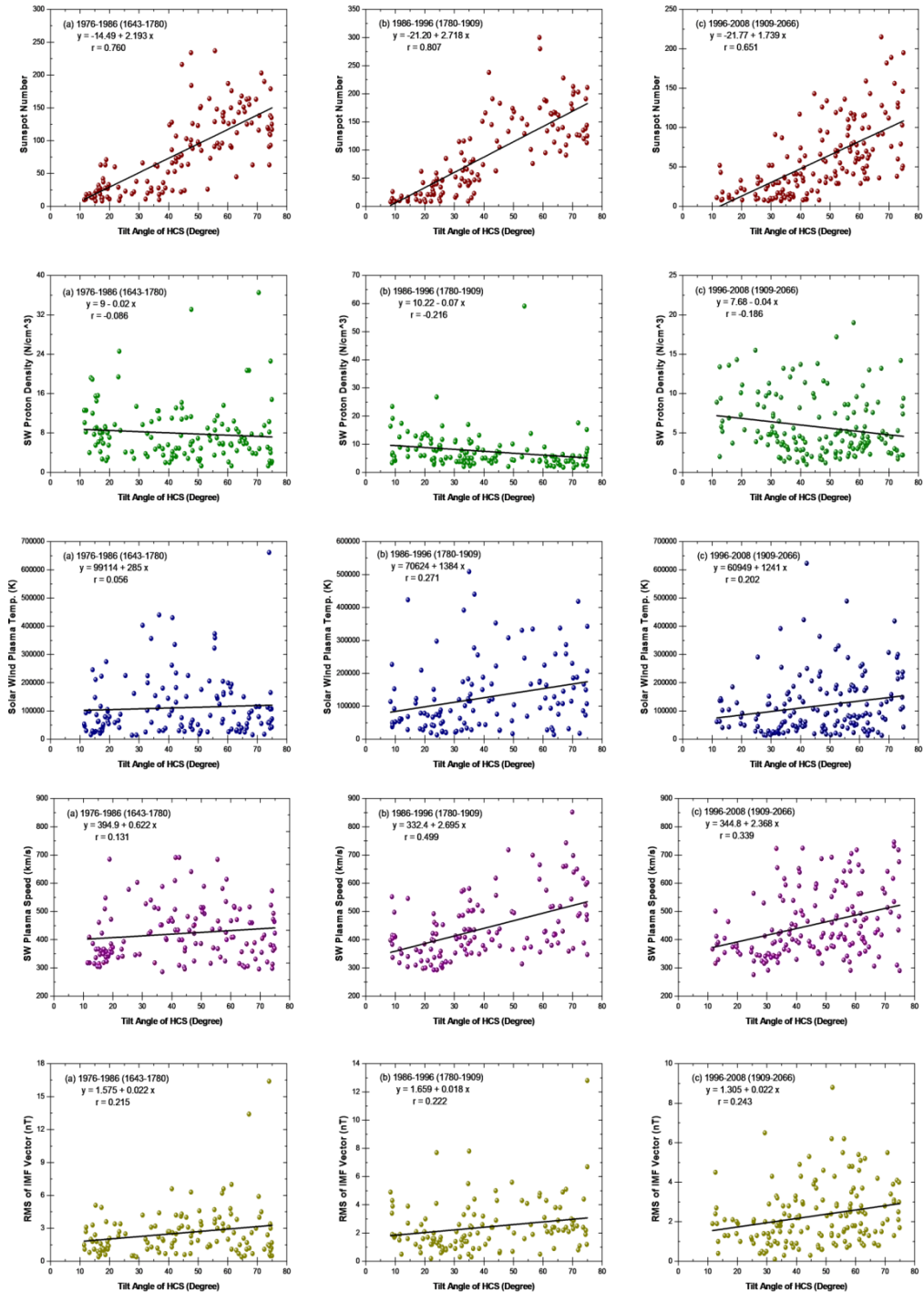

Figure 7: Correlative diagrams of the averages of solar wind speed, sunspot number, ion density, ion temperature and root-mean square of IMF Vector, with respect to tilt angle of HCS for the three considered periods. The correlation coefficients and the best-fit line equations are shown.

A detailed correlative analysis between sunspot numbers (SSN) and tilt angles (TA) has been performed for the solar activity cycles 21, 22 and 23. Figure 7 (second plots) shows the relations between sunspot numbers $R_Z$, and tilt angles of HCS α, during the three phases of solar activity. Second plots (a) to (c) reflect these relations for solar cycles 21, 22, and 23, respectively. High correlation coefficients with nearly similar regression lines are obtained for the three considered periods (0.760 for period 1976-86, 0.807 for period 1986-96 and 0.651 for the years 1996-2008). The slope of the fitted line to the 1986-96 data (solar cycle 22), which equals 2.718, was larger than that of the fit to the 1976-86 (the slope is 2.193) and 1996-2008 (the slope is 1.739) data (SAC 21 & 22, respectively). Therefore, the correlation between sunspot number (SSN) and tilt angle (TA) was more pronounced in even solar cycles (22) than in odd solar cycles (21 & 23). It is thus apparent that the tilt angles of current sheet can be used as a meaningful and significant solar parameter, especially in the case of cosmic ray modulations.

We tried to calculate the cross-correlation coefficient of each one of the rest parameters considered here in comparison to the HCS inclinations, for the odd (21 & 23) and even (22) solar activity cycles. Correlative diagrams with respect to tilt angles of HCS for the years 1976-1986, 1986-1996 and 1996-2008 are given in Figure 7 for the heliospheric variables; solar wind proton density (third plots) and plasma temperature (forth plots). A negative correlation between ion density and TA is found, on the contrary, ion temperature and TA relation is positive one. The solar cycle 21 is characterized with correlation coefficient r of -0.086 for ion density and 0.056 for ion temperature versus α, whilst the solar cycle 22 is marked by r = -0.216 for ion density and 0.271 for ion temperature against α, likewise the solar cycle 23 recorded a magnitude of correlation equals -0.186 for ion density and 0.202 for temperature and tilt angle. As we can clearly see, the correlation of both parameters to the HCS tilts is indeed greater during the period 1986-1996 (SAC 22) than during the two periods 1976-86 (SAC 21) and 1996-2008 (SAC 23). The correlation coefficient for each case is indicated in the diagrams. Generally, it is evident that the correlation of the heliospheric parameters; ion density & temperature with tilts of current sheet is better during even cycles than odd cycles. On the other hand, the sensitivities of ion density to changes in tilt angles have the values -0.02, -0.07 and -0.04 for SAC 21, 22 and 23, respectively. In addition to that, the sensitivity of ion temperature to α recorded a greater value of 1384 for SAC 22 than that for SAC 21, which has a value 285, and SAC 23, which has the value 1241.

The last heliospheric parameter; root-mean square of IMF Vector ($\sigma_B$) is studied relative to the HCS inclinations for solar cycles 21, 22 and 23 as shown in Figure 7 (fifth plots). It was observed that there is a somewhat significant positive correlation with nearly similar regression lines in the three considered periods where the correlation coefficients have values of 0.243 in SC 23, 0.222 in SC 22 and 0.215 in SC 21.So the correlation between $\sigma_B$ and α is constant, within experimental errors. Also, the sensitivity of $\sigma_B$ to variations in tilt angles has a similar value equal 0.022 in case of the SACs 21 and 23, while it has a value of 0.018 for the SAC 22. Therefore, the sensitivity of $\sigma_B$ and α is stable, limited and unchanged with α. Thus, it is remarkable that the sigma B and current sheet tilts relation is independent to some extent on the sort of the solar cycle (odd or even) counter to the other four studied solar and heliospheric parameters.

## 4.5 Effects of hysteresis in solar cycle variations between tilt angles of HCS and cosmic ray intensities

We present the hysteresis effect between the tilt angles of HCS and cosmic ray intensities for the past 38 years from May, 1976 to January, 2014 on a solar rotational basis throughout the last complete solar activity cycles 21, 22 and 23. Figures 8-10, which represent the hysteresis phenomenon in the form of ellipses, show the comparison between the modulation loops between cosmic ray intensities recorded at Calgary neutron monitor ($R_O$ = 1.08 GV) (left plots) and Huancayo/Haleakala NM whose cutoff rigidity is 13.45/13.3 GV (right plots), and the tilt angles of the HCS for solar cycles 21, 22 and 23, respectively. Hourly CRI records of NMs were averaged over the periods of 1-solar rotation, 3-rotation, 6-rotation and 12-rotation. Corresponding Carrington rotation's tilt angles of the HCS were used to obtain average angles for the tilt for respective periods. Arrows indicate the direction of time and show that the CR is the leading index in three cycles.

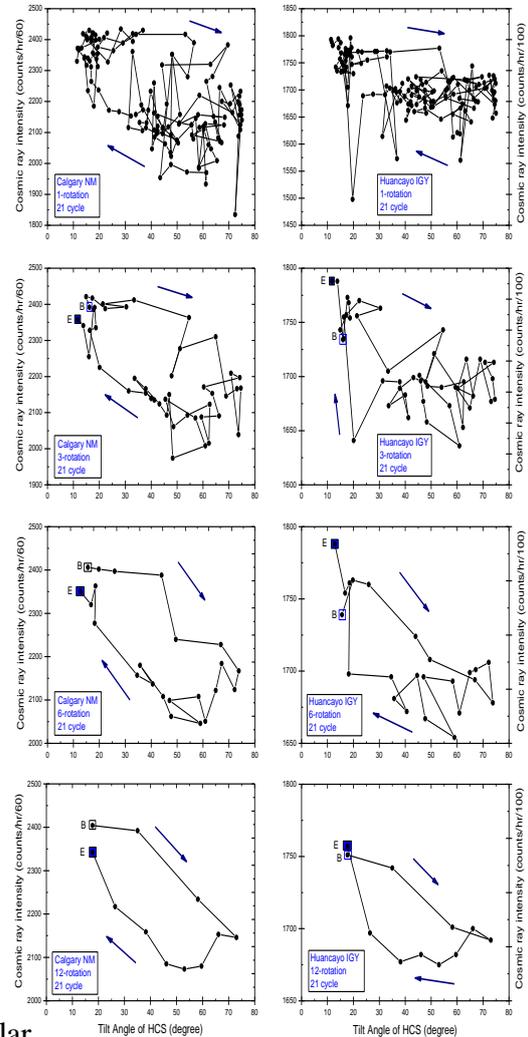

Eight cross-plots for cycle 21 in Figure 8 show a general shape loop indicates a large time lag between CRI and TA during the solar cycle 21. As regards the effectiveness of HCS tilt on the cosmic-ray modulation, during different phases of the activity cycle, the whole modulation loops can be visualized to indicate the followings:

(i) "Inverse-modulation" i.e. decrease in CRI with increase in TA, and vice versa.
(ii) Slow rate of intensity decrease with increasing TA during ascending phase.
(iii) At or nearly the solar maximum and the solar polarity reversal, TA is almost stable with changing CRIs.
(iv) Fast rate of intensity increase with decrease in TA during declining phase.
(v) Intensity decreases at a faster rate with TA as solar activity approaches to minimum level.
(vi) Another interesting observation from the cross-plots shown in Figure 8 is the appearance of "secondary" loops around solar maximum and solar polarity reversal (1979-1980), more clearly seen in 1-rotation and 3-rotation plots, and to some extent in 6-rotation averaged cross-plots.

Figure 8: Hysteresis plots of cosmic ray intensities of Calgary and Huancayo neutron monitors versus HCS Tilt angles, for solar activity cycle 21; data averaged over 1, 3, 6 and 12 Carrington rotations. The B's and E's represent the beginning and the end points of the modulation loops, respectively.

From Figure 8, we have observed that the area of low energy cosmic rays modulation loops at Calgary NM is much larger than that of the high energy cosmic rays loops at Huancayo/Haleakala NM. To quantify this difference, we have estimated the loop areas of solar activity cycle 21 obtained by 1-rotation, 3-rotation, 6-rotation and 12-rotation averaged plot of TA versus CRI, both for Calgary and Huancayo/Haleakala neutron monitor count rates. These estimates are given in Table 3. It is found that the area of the modulation loops of low rigidity Calgary NM is about three times larger than the area of the loops for high rigidity Huancayo/Haleakala NM. It may be return to a great sudden depression in CRI observed in HUN/HAL.

Table 3: Area of CRI modulation loops for SAC 21 for CAL and HUN/HAL NMs.

| Averaged-Solar Rotations No. | Loop Area ($\times 10^5$ counts∗degree/hr) | |
|---|---|---|
| | Calgary | Huancayo/Haleakala |
| 1 | ~ 10.11 | ~ 4.51 |
| 3 | ~ 7.54 | ~ 1.94 |
| 6 | ~ 5.77 | ~ 1.66 |
| 12 | ~ 5.10 | ~ 1.71 |
| 21$^{st}$ SAC | | |

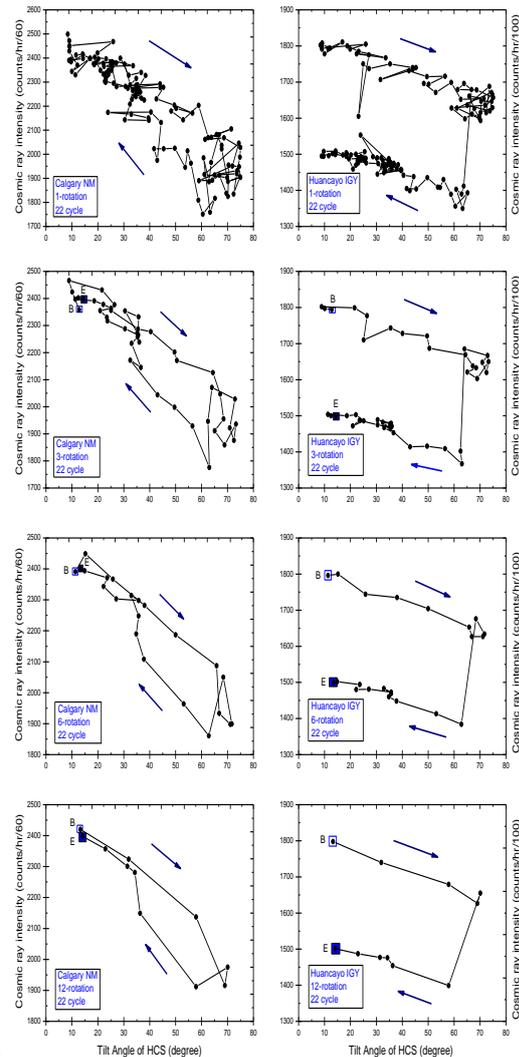

Figure 9 represents the cross-plots between the tilt angle and the CRI for Calgary and Huancayo/Haleakala neutron monitors during the solar cycle 22. These plots correspond to 1, 3, 6 and 12-rotation averaged data. A comparison of cycle 22 modulation loops with that of cycle 21 shows that area of cycle 22 (even cycle) loop is much smaller (about half) than that of cycle 21 (odd cycle) for Calgary NM, indicating a smaller time lag between CRI and TA in cycle 22 than 21, which supports the even–odd asymmetry of the solar cycles. In contrast, the area of cycle 22 (even cycle) loop is much greater (about sevenfold) than that of cycle 21 (odd cycle) for Huancayo/Haleakala NM, indicating a larger time lag between CRI and TA in cycle 22 than 21. Although not as distinct as in cycle 21, some qualitative differences are seen in response of CRI to the TA changes in cycle 22 also. For both NMs in Figure 9, we can say that the rate of CR modulations throughout ascending and descending SA phase is the same. Also, unusual behavior of CR modulations at the SA maximum phase and polar field reversal period in 1990, no correlation is observed between α and CR modulation, showing an invariant in the rate of CR modulation at nearly $\alpha \cong 62°$. Furthermore, secondary small loop around the same period is another feature of these plots.

Figure 9: Same as Figure 8 for solar activity cycle 22.

Table 4 shows the estimated loop areas of solar cycle 22 from plot of TA versus CRI for Calgary and Huancayo/Haleakala neutron monitors. Unlike the SAC 21, the area of the modulation loops of Huancayo/Haleakala NM is almost three and a half times the loop area of Calgary NM.

**Table 4: Area of CRI modulation loops for SAC 22 for CAL and HUN/HAL NMs.**

| Averaged-Solar Rotations No. | Loop Area ($\times 10^5$ counts∗degree/hr) | |
|---|---|---|
| | Calgary | Huancayo/Haleakala |
| 1 | ~ 4.40 | ~ 13.85 |
| 3 | ~ 4.25 | ~ 13.79 |
| 6 | ~ 3.91 | ~ 14.62 |
| 12 | ~ 3.28 | ~ 13.77 |
| 22$^{nd}$ SAC | | |

Cross-plots between CRI and TA for cycle 23 are shown in Figure 10. A slow rate of decrease in CRI with increase in TA during the increasing phase, a larger intensity decrease without any significant change in TA (almost no relation between CRI and TA) at or nearly close to solar maximum and polarity reversal, a fast increase in CRI with decrease in TA during decreasing phase of solar cycle 23 is apparent from Figure 10. These observations, from the cross-plots between CRI and TA for solar cycle 23, are almost similar to those observed in cycle 21 during different phases of the solar cycles; a weak CRI response to TA changes during the increasing phase, almost no-response in and around solar maximum and polarity-reversal period, and a stronger response during decreasing phase.

Similar to modulation loops of the SAC 21, the area estimates which given in Table 5 show that the area of the modulation loops of low rigidity Calgary NM is about three times larger than the area of the loops for high rigidity Huancayo/Haleakala NM.

In order to support our obtained results, we have used CRI data of two neutron monitors of low (Calgary, $R_O = 1.08$ GV) and high cutoff rigidities

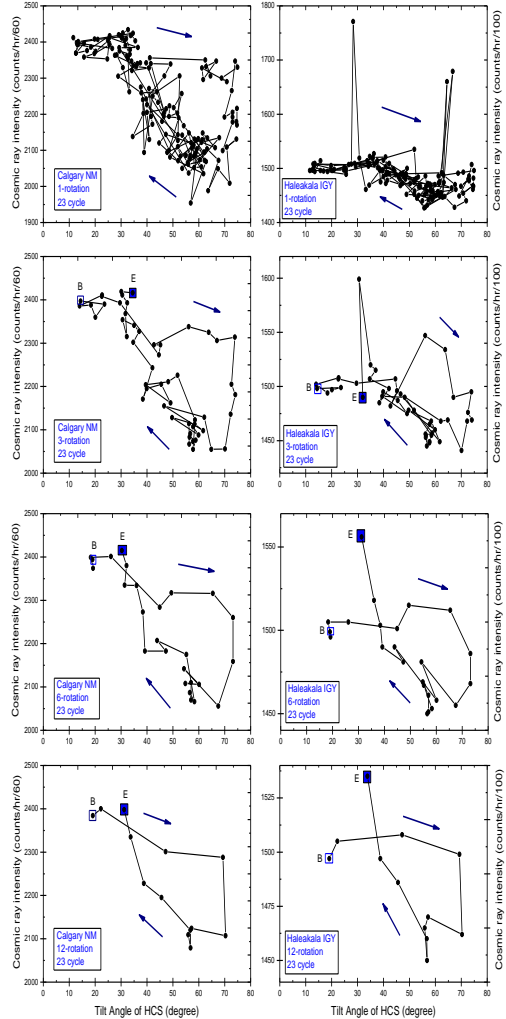

Figure 10: Same as Figure 8 for solar activity cycle 23.

(Huancayo/Haleakala, $R_O$ = 13.45/13.3 GV). It is clearly seen that these plots with Huancayo/Haleakala NM data are different to those obtained with Calgary NM data in respective solar cycles due to different particles energies. Badruddin et al. (2007) have estimated the loop area of individual solar cycles from six-rotation averaged plot of TA versus CRI, for both Climax and Oulu neutron monitor count rates. He found that the area of the modulation loops of both the odd cycle 21 is about three times larger than the area of the loop for even cycle 22. However, cycle 23 area is not as larger as 21; an indication from the data up to 2006.

Table 6: Area of CRI modulation loops for SAC 23 for CAL and HUN/HAL NMs.

| Averaged-Solar Rotations No. | Loop Area ($\times 10^5$ counts∗degree/hr) | |
| --- | --- | --- |
| | Calgary | Haleakala/Huancayo |
| 1 | ~ 2.12 | ~ 1.58 |
| 3 | ~ 3.17 | ~ 1.29 |
| 6 | ~ 3.41 | ~ 0.76 |
| 12 | ~ 2.89 | ~ 0.72 |
| $23^{rd}$ SAC | | |

## 4.6 Time-Lag between CRIs and HCS tilts in solar cycles 21, 22 and 23

The correlation coefficient between cosmic ray intensity and tilt angles of HCS with lag time has also been calculated for the three solar activity cycles using the method of "minimizing correlation coefficient method". Here we have selected both the series CRI and TA for the same period with zero lag time and then shifted one series by a step of one Carrington rotation and calculated the cross correlation coefficient between both the series. Similarly, the other series has also been shifted by one Carrington rotation and the new value of cross correlation coefficient is calculated. As such, the time (number of shifted Carrington rotations) is obtained, when the anti-correlation coefficient is maximum. This is the lag time between both the series CRI and TA.

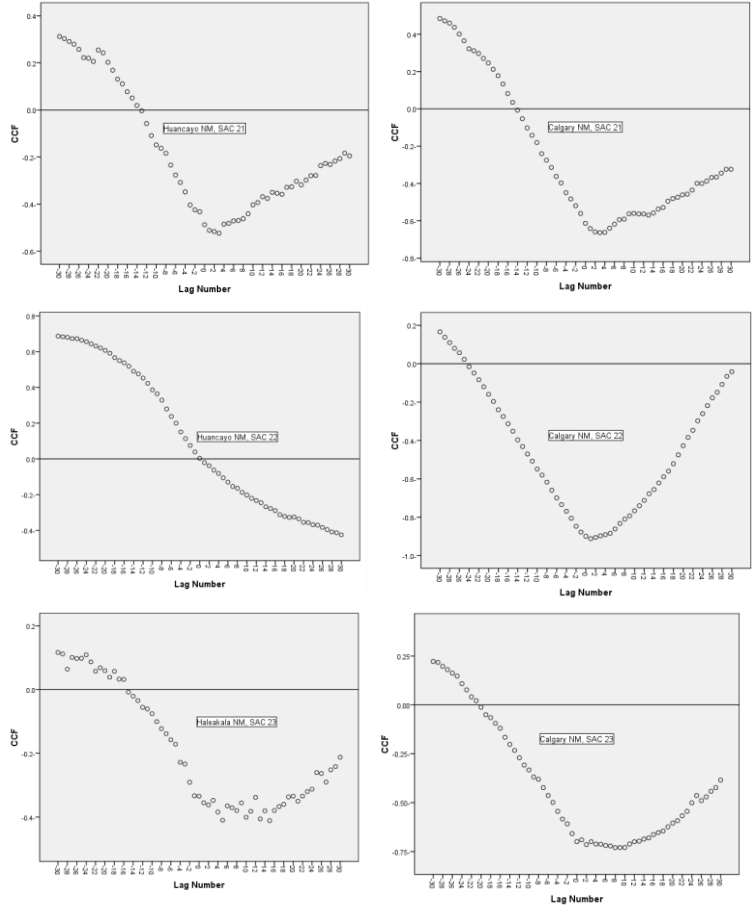

Figure 11: Cross correlation coefficient factors between tilt angle of HCS and GCR intensity (Calgary and Huancayo/Haleakala NMs) with different time lags, for solar activity cycles 21, 22 and 23.

The lag time (in Carrington rotations) versus the cross correlation factor (CCF) graphs for the three solar cycles are plotted in Figure 11. From the optimum value of the correlation coefficient (at the time of maximum anti-correlation coefficient), it is inferred that the lag time ~ 3 Carrington rotations during the odd solar cycle 21 for both NMs (see also Table 7). On the other hand, the lag time during even cycle 22 is found to be 1 Carrington rotation for CAL NM and to be 30 for HUN NM. The lag time between tilt angle and CRI during other odd cycle 23 is again found to be (8-15 Carrington rotations). It may be worth mentioning that the lag time is larger, in odd cycles 21 and 23 for Calgary (NM of low rigidity), when the recovery phase of the ~ 11 year modulation cycle lies in qA < 0 epochs; in this polarity state, positively charged particles enter the heliosphere via the HCS. The lag time is much smaller in even solar cycle 22, when the recovery phase of the ~ 11 year modulation cycle lies in qA > 0 polarity state and the access of cosmic ray particles to the inner heliosphere is through Solar Polar Regions. In contrast, the lag time is larger, in even cycle 22 for HUN/HAL (NMs of high rigidities), than that in odd cycles 21 and 23.

Gupta et al. (2006) have calculated the cross correlation coefficient between CRI, observed at Kiel ($R_C$ ~ 3 GV), and TA by shifting of both the series one by one by a step of one month. It is observed that during odd cycles 21 and 23 the time lag between CRI and TA is ~17 and 11 months at the time of maximum anti-correlation coefficient (r ~ −0.8) whereas for even cycle 22 the time lag has been found to be ~ 2 months at the time of maximum anti-correlation coefficient (r ~ −0.9).

Badruddin et al. (2007) determined the time lag between CRIs (observed at Climax NM, whose $R_O$ = 2.97 GV, and Oulu NM, whose $R_O$ = 0.77 GV) and TA in solar cycles 21, 22 and 23; the correlation coefficients have been calculated between the TA of HCS and Carrington rotation averages of CRI, by introducing successively time lags of 0 to 29 Carrington rotations. From the optimum value of the correlation coefficient, it is inferred that the time lag is 24 Carrington rotations during the odd solar cycle 21. On the other hand, the time lag during even cycle 22 is much smaller (1–2 Carrington rotations). The time lag between TA and CRI during other odd cycle 23 is again found to be large (12–14 Carrington rotations).

Table 7: Correlation coefficient between CRI and TA with time lag for solar cycles 21, 22 and 23

| SAC | Station Name | Lag Time (Carrington rotations) | Maximum value of CCF |
|---|---|---|---|
| 21 | Calgary | 3 | -0.664 |
|  | Huancayo | 3 | -0.524 |
| 22 | Calgary | 1 | -0.912 |
|  | Huancayo | 30 | -0.425 |
| 23 | Calgary | 8 | -0.730 |
|  | Haleakala | 15 | -0.411 |

## 4.7 Running Cross Correlation Coefficients between HCS Tilt, CRI and Solar/Heliospheric Parameters

To support the time lag findings, the running cross correlation method has been used to study the relationship between tilt angle of HCS and, CRI observed at Oulu NM, Sunspot number ($R_Z$), solar wind speed (SWS) and ion density (n). Figure 12 (panels a to d) shows the linear running cross correlation coefficient c(t) between Alpha & CRI at Oulu (Fig. 12a), Alpha & $R_Z$ (Fig. 12b), Alpha & SWS (Fig. 12c) and Alpha & n (Fig. 12d) from 6/1976 to 1/2012 for CNs from 1656 to 2104. In the upper panel, we denoted the epochs of minimum solar activity by min `s (with up arrows) while the reversals of the Sun`s polar magnetic field (within epochs of maximum SA) are expressed by vertical shaded area. In this analysis, we used a time window of width T centered at time t: [t − T/2, t + T/2]. The cross correlation coefficient c (t) is calculated for data within this window. Then the window is shifted in time by a small time step ∆t < T and the new value of the cross correlation coefficient is calculated (El-Borie, 2001; Meera Gupta et al., 2006). We have used the time shifting of one CN to calculate the correlation coefficient for each CN between TA - CRI, TA - $R_Z$, TA - SWS and TA - n for the period 1976 to 2012. The time window of 30 CNs has been taken. This value was chosen to match two contradictory requirements (El-Borie, 2001);

(i) Uncertainty of the calculated c (t) is smaller for large T.

(ii) T should be small in order to reveal the fine temporal structure of the cross correlation function. The values |c (t)| = 1 represent the total coincidence of the data sets. C(t) = 0, no coincidence, and the sign of c(t) is the same as the sign of a slope of sensitivity relation between data sets.

From Figure 12 (plots a to d), we concluded the followings:

1) There is a high negative correlation between (TA, CRI) and a low negative correlation between (TA, n), while a high positive correlation is found between (TA, $R_Z$) and a moderate positive correlation for (TA, SWS).
2) Sometimes there is no correlation (c (t) ≈ 0) between alpha and the three solar parameters.
3) The cross correlation coefficients between α and CRI reached maximum values leading to positive peaks near the years of IMF polarity reversal and maximum solar activity (1979/80, 1989/90 and 1999/2000) confirmed the results reported before (El-Borie, 1999). Also, c (t) of alpha and n recorded positive peaks over the epochs of the same years. In contrast, around the years of reversal of IMF polarity (maximum solar activity), negative peaks are emerged in case of (alpha and $R_Z$) and (alpha and SWS).
4) Around years of minimum solar activity (1986/87, 1996/97 and 2008/09), the magnitude of running correlation coefficient recorded a high magnitude (|c (t)| ≈ 0.6-0.7) between TA and CRI beside the case of TA and $R_Z$ over the years 1986/87. While the cross correlation coefficient showed a moderate value (|c (t)| ≈ 0.2-0.5) for the

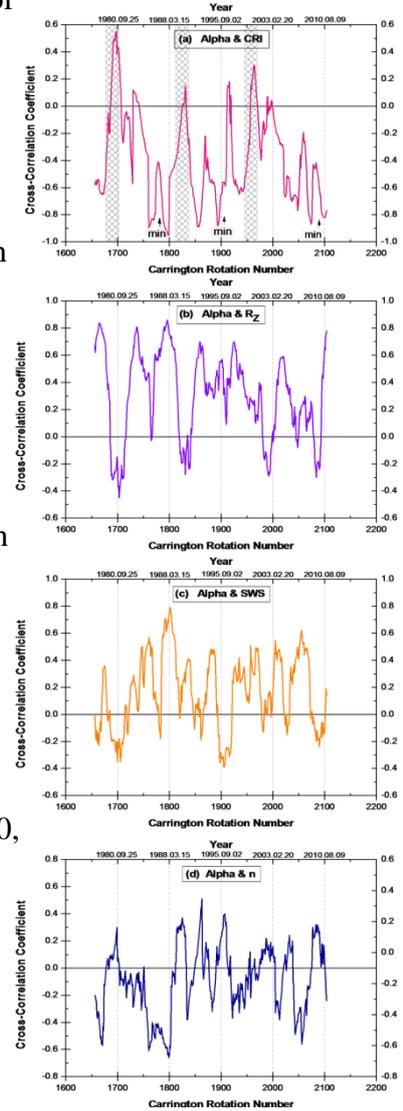

Figure 12: Linear cross-correlation coefficients between the cosmic ray intensity (CRI) observed at Oulu NM, sunspot number ($R_Z$), the solar wind speed (SWS) and the solar wind plasma density (n) respectively, with tilt angles of HCS throughout the period June 1976-January 2012 (from CN 1656 to CN 2104).

rest years of minimum SA of alpha and SSN and for the two data sets (alpha & SWS) and (alpha & n).

5) There are quasi-periodicities (the peak-to-peak locations) in the observed correlation coefficients where the data series (α & CRI) has an average periodicity of 44 CN and an average periodicity of 54 CN, 43 CN and 48 CN for the data sets (α & $R_Z$), (α & SWS) and (α & n) are noted, respectively.

6) It is also evident that there is a 4.5-year periodicity in the observed peaks occurred on the average which represents a round half solar cycle.

7) The variability of correlation waves in the four charts (a-d) are different for the three intervals of periodically opposite magnetic polarities of IMF (1981/89, 1991/99 and 2001/2009). Higher correlations are observed for qA < 0 (1981-1989).

8) The TA behavior is similar during the rising phases of the solar cycles 21, 22 and 23. In contrast, it is different during the declining phase of the solar cycle 23 than the solar cycles 21 and 22 (Fig. 12a and b). The similarities in the TA evolution during the rising of cycles 21 and 22 have also been reported (Suess et al., 1993; Cliver, 1993; Cliver & Ling, 2001).

9) Finally, there is a time lag of about 4 months between the peak to peak of the TA and CRI correlation, as well as the peak to peak of the SSN and TA correlation. The temporal variations in the size of the polar holes correspond reasonably well to the variations in the CRIs observed at Earth, with time lags of about 4 months. Since the coronal holes play a key role in determining the spatial structure of the IMF, which essentially determines the location and configuration of HCS on the source surface.

## 5. Conclusions

The heliosphere is known as a Sun influential region in the interstellar medium. The heliospheric current sheet separates the two oppositely directed magnetic polarity hemispheres of the heliosphere. Thus, the neutral sheet tilts are expected to play a significant and remarkable role in the study of solar and magnetic disturbances in the heliosphere and on the modulations of energy particles. In the present work, we have presented a systematic correlation analysis to investigate the dependence of cosmic ray intensities observed at sixteen NMs for the 1976-2014 period (the median rigidities of response are ranging from 16 to 33 GV) and solar/interplanetary parameters, on the waviness of heliospheric current sheet for different epochs of IMF. Our conclusions can be summarized as follows:

1) The solar activity cycle (≈11 years) variation of HCS tilt angles has been emphasized. The neutral sheet is highly inclined around solar maximum years (during epochs of IMF polarity reversal) and dips to lower inclinations (or nearly equatorial) near SA minimum. The current sheet inclination angles exhibited well with the sunspot numbers and inversely with CRIs through the considered period (May 1976 – January 2014).

2) For a long term study of cosmic ray intensity, it was found that CRI varies periodically as α every about 11 years (SAC). Also, there were some large decreases in intensity occurred in 1982, 1984, 1986, 1991 and 2001. These decreases were not related to any changes in the inclinations of current sheet, α. Other depressions in the count rates of CR occurred in 1978, 1979, 1989, 1998 and 2000, following relatively large excursions in α.
3) The time profiles of solar/interplanetary parameters (solar wind speed SWS, solar wind temperature T, ion density n and sigma of IMF vector $\sigma_B$) show a weak response to changes in the inclinations of HCS during the cycles 21, 22 and 23. These parameters changes very slowly with time and have a weak dependence at the minimum of the solar cycle 23.
4) There is a negative and high correlation between tilt angles of HCS and the CR intensities. The magnitude of the correlation depends on the solar magnetic polarity states, whereas when the solar magnetic field pointed inwards (negative polarity state), the regression coefficient r have a value of 0.85 for the period 1981/88 (0.91 for 2001/08) and a decrease in the intensity of ≈ 7 GeV cosmic ray at Potch. Neutron Monitor was 0.175 %/deg (0.246%/deg), respectively. Whilst during the period of solar magnetic pointed outward (1991/98 epoch), r was 0.73 and the sensitivity was 0.161%/deg. This lower sensitivity for positive field is in consistent with the gradient drift model and with the change from a flat cosmic ray maximum (square wave) to a peaked maximum as a triangle wave (Smith, 1990; Cliver, 1993). The drift theory describing the latitudinal cosmic ray gradient on either side of the neutral sheet should be opposite for the opposite solar magnetic phases of the heliosphere.
5) The rigidity-dependence of the average sensitivity was found to be changed according to the solar polar magnetic field states. During qA < 0, it is represented by a power law, $R^{-1.00}$ for the period 1981-88 and $R^{-1.48}$ for the interval 2001-08; whilst it was $R^{-1.35}$ for 1991/98 epoch (qA > 0), indicating that the cosmic ray modulations were harder in years after the reversal IMF 1999/2000 than before it.
6) The form of the average cosmic ray modulation spectrum with the HCS tilt angles varies as a function of IMF polarity switch every ≈11 years (as well as, every 22 years). It was also found that larger CR modulation for lower particle energies. These results have been discussed according to the drift theory (J.R. Jokipii et al., 1981; 1983). Furthermore, the rates of sensitivities for different particle energies caused by HCS tilts are higher during qA < 0 solar minimum periods than the corresponding years in the qA > 0 epoch, for the 22 magnetic solar cycle 1991-2012.
7) Positive and moderate correlations are observed between α and the average of SWS during the last three complete solar cycles 21, 22 and 23, and it is dependent on the essence of each SA Cycle. Whereas for the even solar cycle (22), the degree of the correlation is much steeper than that of the odd solar cycles (21 & 23).
8) $R_Z$ has a strong correlation with neutral sheet tilts. Such as SWS, the correlation between SSN and TA is more pronounced in even solar cycle (22) than in odd SC (21 & 23).
9) A weak and negative relation is found between SW density and TA; on the other hand, inclinations of HCS are linked with SW temperature through weak positive correlations. It is remarkable that the sigma B and alpha relation is not dependent on the type of the solar cycle whether it is odd or even which contradict the other four studied of solar and interplanetary indices.
10) During odd cycles, increasing phase of solar activity cycle lies in qA > 0 state when particle access to the inner heliosphere is via polar regions. As the odd-numbered cycle progresses, intensity decreases at a slow rate during qA > 0; in this epoch, the current sheet is less effective since positively charged particles enter the heliosphere through polar regions. When the solar activity becomes substantially high, the heliosphere is likely to be filled up with interaction regions extending to large extent in latitude and longitude. These regions will act as barriers for the particles and thus efficiently reducing the cosmic ray intensity when the solar activity is high. Around solar maximum when the tilt of the HCS is close to 90°, the charged particles will come across the fields of both positive and negative polarities in polar regions and they drift sometimes inward and sometimes outward (Zhang, 2003). In such condition, tilt angle of the HCS is unlikely to play an effective role in modulation. After the polarity reversal around solar maximum, the state of the heliosphere changes to qA < 0 when particle access is along the current sheet. Although presence of interaction regions around the solar maximum may not be conducive for drifts, their limited presence, and drifts along the current sheet, probably accelerates the rate of decrease in intensity. Furthermore, as the solar activity decreases toward minimum, the tilt of the HCS decreases and the intensity recovers.

11) In the even-numbered cycles, the early phase of intensity decrease lies during qA < 0 polarity state. As the cycle progresses, tilt angle gradually increases and the HCS being more effective due to particle access along the current sheet, intensity decreases at a faster rate. Around solar maximum additional effect of diffusive barriers further reduce the intensity. At solar maximum when the current sheet tilt is close to 90° drift is sometimes inwards and sometimes outwards and diffusive barriers of interaction regions appears to be the main agents responsible for cosmic ray modulation. After polarity reversal around solar maximum, the heliosphere changes to qA > 0 state when the particle access is via polar regions. Particle access via solar poles, together with diffusive barriers being less effective in this situation and producing a quick intensity increase (recovery) after solar maximum without much time lag.

12) Finally, so far as the response of TA change to CRI variation is concerned i.e. from modulation point of view, current solar activity cycle (23) is similar to earlier odd cycle 21, but evolutions of cycle 23 and 22 appear quite different from each other.

13) The lag time between tilt angle of HCS and CRI is found to be larger in odd cycles 21 and 23 than even cycle 22 for Calgary (NM of low rigidity), while the lag time is lager, in cycle 22 for HUN/HAL (NMs of high rigidities), than that in cycles 21 and 23.

14) The technique of running cross-correlation function was used to examine the observed variations in CRIs, which changed by the tilts of HCS, and explore the relations between TA and solar/interplanetary parameters. Near the years of maximum SA, there were remarkable positive correlations leading to positive peaks for the two sets (alpha & CRI) and (alpha & n), and negative peaks in case of the sets (alpha & $R_Z$) and (alpha & SWS). In addition, there are quasi-periodicities (on average ≈ 48 CRs) in the correlations between α and the other four indices. It is also evident that there is about 4.5 year (half of the solar cycle period) periodicity in the observed peaks. The observed magnitudes of positive or negative correlations for α, implies that the HCS tilt angle is well solar parameter.

NOTATION

The interdisciplinary nature of this work introduces a great many acronyms and notations that are in common use in any one field but may not be so well known by scientists from another field. We therefore list them here.

Acronyms

| | |
|---|---|
| IMF | Interplanetary Magnetic Field |
| HCS | Heliospheric Current Sheet |
| CRs | Cosmic Rays |
| CRI | Cosmic ray Intensity |
| Rz | Sunspot number |
| TA | Tilt Angle |
| SWS | Solar Wind Speed |
| n | Ion Density |
| GV | Giga Volt |
| qA> 0 | IMF points outward and away from the Sun |
| qA<0 | IMF points inward and toward the Sun |
| α | Heliospheric Current Sheet Tilt |
| SA | Solar Activity |
| SAC | Solar Activity Cycle |
| CN | Carrington rotation Number |
| Rc | Cutoff Particle Rigidity |
| Rm | Median primary Rigidity |
| AU | Astronomical Unit |
| GCR | Galactic Cosmic Ray |
| SCR | Solar Cosmic Ray |
| NM | Neutron Monitor |
| WSO | Wilcox Solar Observatory |
| $R_S$ | Solar Radii |
| Fds | Forbush Decreases |
| m.a.s.l | meters above sea level |